\documentclass[english]{cfm2007}

\begin{document}

\cfmtitre{Direct Numerical Simulation of structural vacillation in the
  transition to geostrophic turbulence}

\cfmauteur{Anthony Randriamampianina$^1$, Pierre Maubert$^1$, Wolf G. Fr\"uh$^2$ \& Peter
  L. Read$^3$}

\cfmadresse{$^1$ Institut de Recherche sur les Ph\'enom\`enes Hors Equilibre, 
  UMR 6594 CNRS \\ Technop\^ole de Ch\^ateau-Gombert. 49, rue Fr\'ed\'eric
  Joliot-Curie. BP 146 \\ 13384 Marseille cedex France \\ $^2$ School of
  Engineering and Physical Sciences, Heriot Watt University \\ Riccarton, Edinburgh, EH14 4AS, UK
  \\ $^3$ Atmospheric, Oceanic and Planetary Physics \\ University of Oxford,
  Department of Physics \\ Clarendon Laboratory, Parks Road, Oxford, OX1 3PU, UK
   \\ Corresponding author: randria@irphe.univ-mrs.fr}

\cfmabstract{The onset of small-scale fluctuations around a steady convection
 pattern in a rotating baroclinic annulus filled with air is investigated
 using Direct Numerical Simulation. In previous laboratory experiments
 of baroclinic waves, such fluctuations have been associated with a flow
 regime termed \textit{Structural Vacillation} which is regarded as the first
 step in the transition to fully-developed geostrophic turbulence.
          }

\cfmresume{Le d\'eveloppement de fluctuations de petite \'echelle dans un
          \'ecoulement convectif permanent d'air en cavit\'e tournante barocline est analys\'e par simulation
          num\'erique directe. Des travaux exp\'erimentaux ant\'erieurs ont
          associ\'e de telles fluctuations au r\'egime de vacillation
          structurelle, consid\'er\'e comme la premi\`ere \'etape vers la
          turbulence g\'eostrophique.}

\cfmkeywords{Structural Vacillation ; Barotropic instability ; Geostrophic turbulence}

\section{Introduction}

The transition to disordered behaviour in the form of `Baroclinic Chaos'
 provides an important prototypical
 form of chaotic transition in fluid dynamics. This is of particular
 geophysical relevance in the context of understanding the origins of chaotic
 behaviour and limited predictability in the large-scale atmospheres of the
 Earth and other terrestrial planets, such as Mars, and in the oceans
 (e.g~\cite{Pierrehumbert95,Read98,Read01}). For many years, the differentially-heated,
 rotating cylindrical annulus has proved a fruitful
 means of studying fully-developed, nonlinear baroclinic instability in the
 laboratory (\cite{Fowlis65},~\cite{Fruh97}). Transitions within the regular wave regime
 follow canonical bifurcations to low-dimensional chaos, but disordered flow
 appears to emerge via a different mechanism involving small-scale secondary instabilities.  
Not only is the transition to geostrophic turbulence less well understood than
 those within regular waves, but also the classification and terminology for
 the weakly turbulent flows is rather vague. 
 Various terms applied include Structural Vacillation or Tilted-Trough
 Vacillation which both refer to fluctuations by which the wave pattern
 appears to change its orientation or structure in a roughly periodic fashion.
 These fluctuations have been explained by the growth of  higher order radial
 mode baroclinic waves (\cite{Weng87}), by barotropic instabilities,
 or by small-scale secondary instabilities or eddies, which lead to erratic
 modulations of the large-scale pattern (\cite{Hignett85}).  
Subsequent development within this so-called `transition zone'  leads to the
 gradual and progressive breakdown of the initially regular wave pattern into
 an increasingly disordered flow, ultimately leading to a form of
 'geostrophic turbulence' (\cite{Read01}).

\section{The numerical model}

 The physical model comprises a body of air contained between two vertical,
 coaxial cylinders held at constant temperatures and two horizontal insulating
 rigid endplates separated by a distance of $d$. 
  The inner cylinder at radius $r = a$ is cooled ($T_a$) and the outer
 cylinder at radius $r = b$ is heated ($\Delta T = T_b - T_a = 30K$).
 The whole cavity rotates at a uniform rate around the axis of the cylinders. 
The geometry is defined by an aspect ratio, $A = d/(b-a)=3.94$ and a curvature
 parameter, $R_c = (b + a)/(b-a)= 3.7$, corresponding to the configuration used by \cite{Fowlis65}
in their experiments with liquids. 
Following a long-established tradition (e.g.~\cite{Fowlis65}), 
 the analysis of the flow in a fixed geometry for a given fluid, here air at
 ambient temperature with $Pr=0.707$, is made by
 varying two main parameters:
 the Taylor number $Ta$ and the thermal Rossby or `Hide' number $\Theta$,
\begin{displaymath}
    Ta=\frac{4\Omega^2 (b-a)^5 }{\nu^2 d} ,\,\,\,\,\,
\Theta= \frac{gd\alpha\Delta T}{\Omega^2(b-a)^2}.
\end{displaymath}
The Navier-Stokes equations coupled with the energy equation via the
 Boussinesq approximation are solved using a pseudo-spectral collocation-Chebyshev
 Fourier method associated with a second order time scheme.

\section{The onset of SV regime}

\begin{figure}
  \begin{center}
\includegraphics[width=0.52\textwidth]{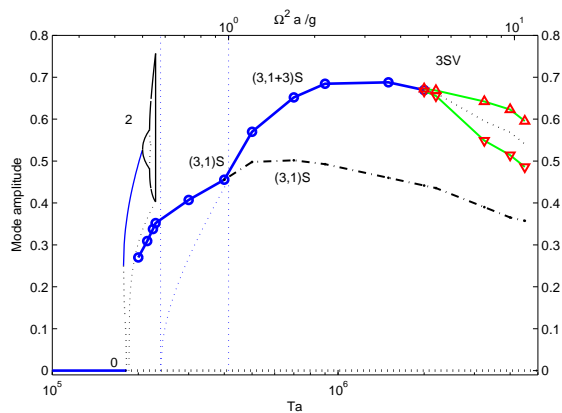}
\caption{\label{Bifurcations}Summary of the regimes obtained by the numerical
  model for a temperature difference of $\Delta T=30\rm K$.
 The axis at the bottom shows the Taylor number while the axis at
 the top shows the ratio of the centrifugal term to gravity at the inner radius.}
\end{center}
\end{figure}

Figure~\ref{Bifurcations} shows the equilibrated amplitude of the dominant
 azimuthal mode against the Taylor number.  For steady solutions, the
 equilibrated amplitude of that wave mode is shown while for time-dependent
 solutions, the maximum and minimum as well as the mean amplitude are given.
  As such, the figure is a bifurcation diagram representing  the sequence
 of three basic solutions, the axisymmetric solution on the $x$-axis and the
 $m=2$ and $m=3$ solutions projected onto the same plane. 
 The instability of the axisymmetric solution and the $m=2$ solution branch
 are reproduced from the earlier study in \cite{Randria06}. The present study
 focusses on evolution of the $m=3$ solution branch when increasing
 progressively the rotation rate up to $Ta=5. \times 10^6$.
  A transition in the flow structure is observed, where 
 the three-dimensional structure of the flow responds to a shift
 in the balance between gravity and the centrifugal force,
 quantified by the local Froude number, $Fr = \Omega^2 r /g$,
 and the dominant physical process changes from baroclinic
 instability to convection due to radial buoyancy. 
 The transition of this convection to chaotic behaviour
 is fundamentally different from that observed in the transition
 to the chaotic flow observed at lower rotation rates. 
 Rather than via a sequence of low-dimensional, quasi-periodic states,
 the large-scale convection developed small-scale instabilities,
 which have been previously suggested as the origin of
 Structural Vacillation (SV) within the transition to geostrophic turbulence.

\begin{figure}
 $(i)$~\hspace*{0.5\textwidth}~$(ii)$\hfill~\\
\includegraphics[width= 0.48 \textwidth]{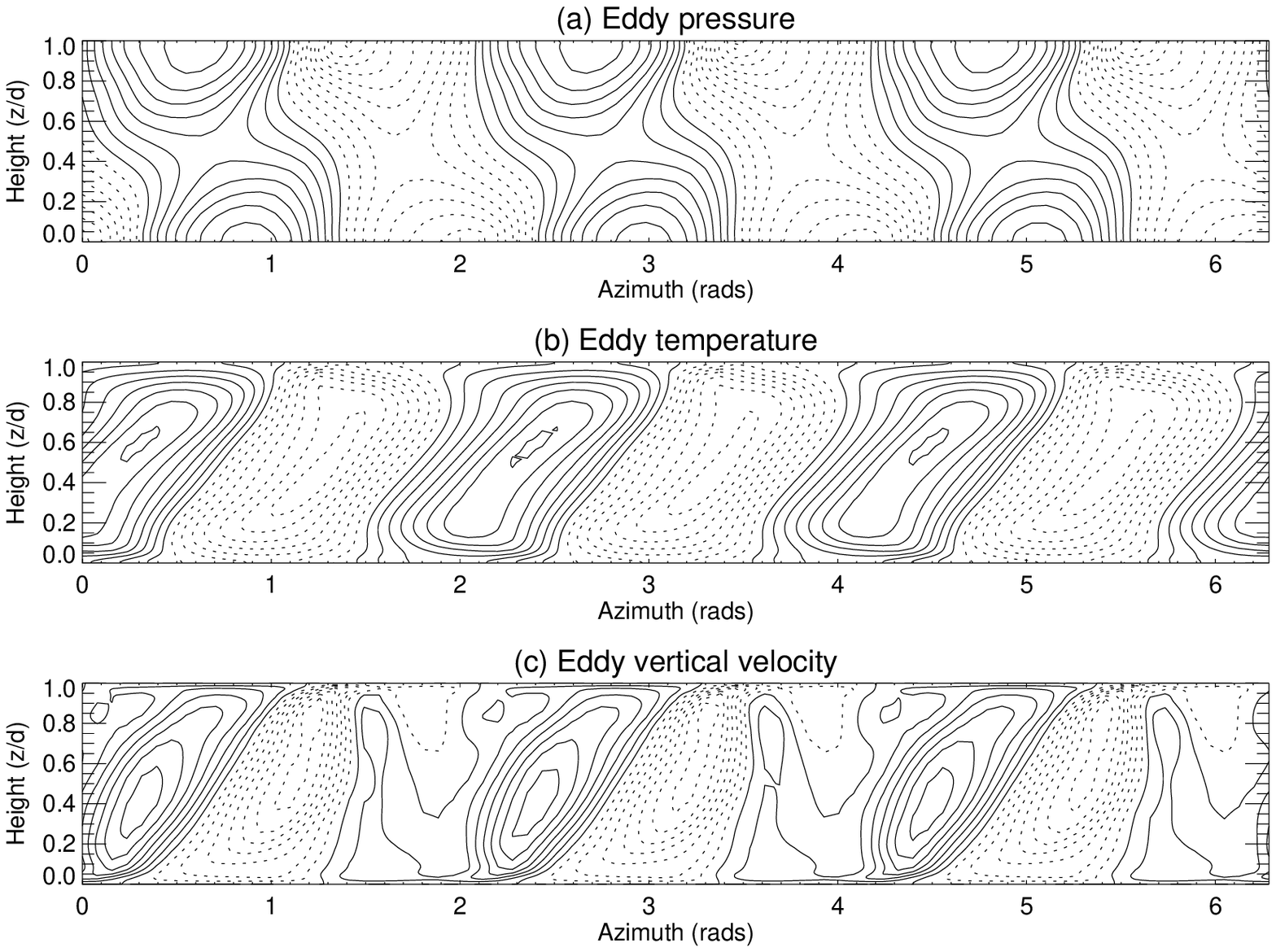} \hfill
 \includegraphics[width= 0.48\textwidth]{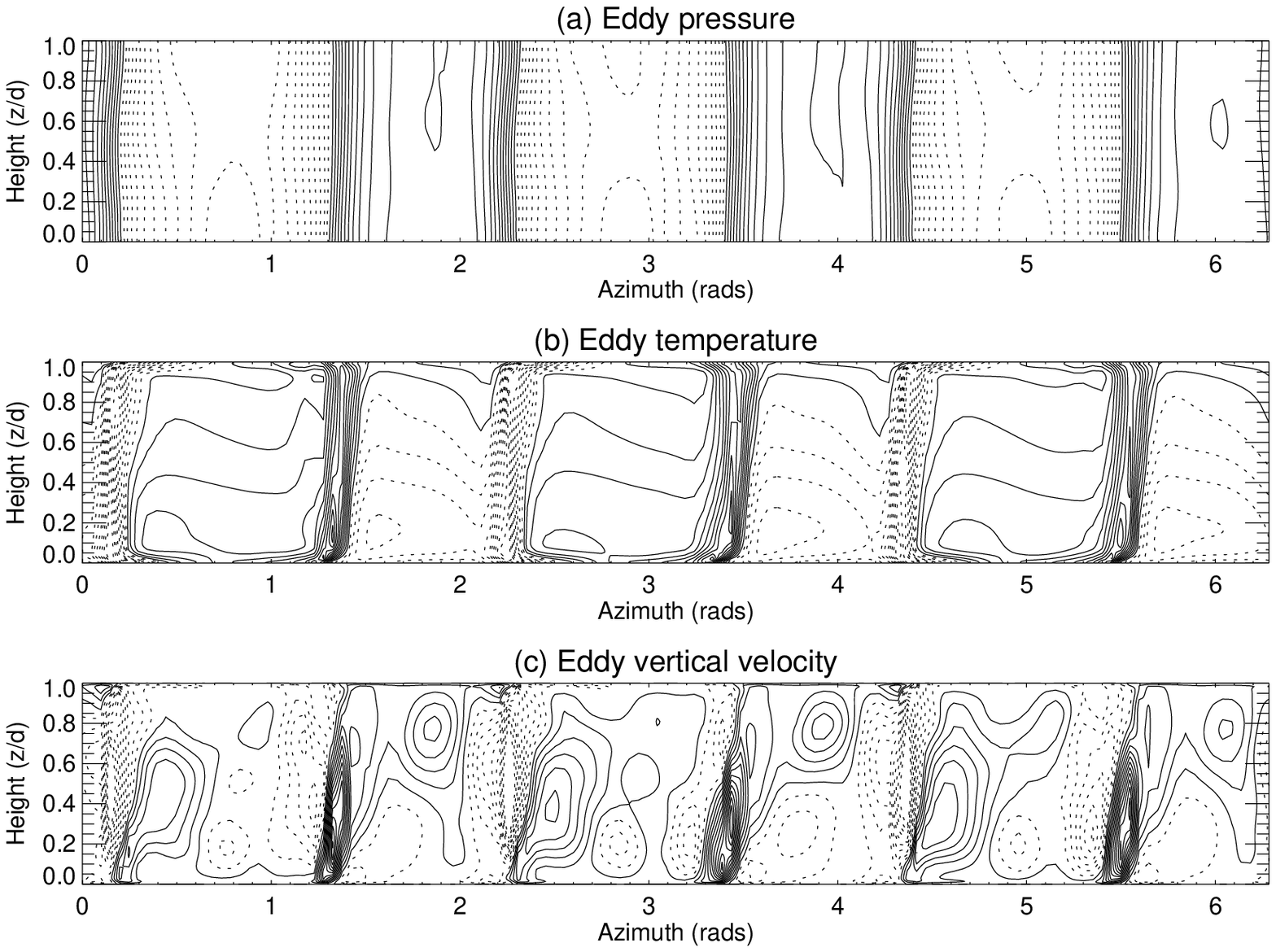}
\caption{\label{TZfields1}Azimuth-height maps of eddy fields (with azimuthal
  flow removed) of a wave-3 simulation at $(i)$ $Ta= 0.235 \times 10^6$ ;
  $(ii)$  $Ta= 4. \times 10^6$ . (a) Pressure (contour interval = 0.25
  dimensionless units), (b) Temperature (contour interval = 0.05,
 normalised to an imposed temperature difference of $\pm 1.0$)
 and (c) Vertical velocity (contour interval = 0.5 dimensionless units). }
\end{figure}

 The Froude number taken at the inner radius $Fr_a \equiv \Omega^2 a /g$
 was used as the secondary axis on top of the graph
 in Figure~\ref{Bifurcations} and the location of $Fr_a=1$
 is highlighted by the dotted vertical line at $Ta=414,000$. 
 The first vertical line, at $Ta= 239,000$, indicates where the centrifugal
 force equals gravity at the outer cylinder, $Fr_b=1$. 
 Below that Froude number or Taylor number,
 gravity is the stronger of the two forces everywhere in the annulus. 
 It can be seen that $Fr_a = 1$ corresponds to a place on the solution
 branch where the radial structure of the steady wave changes.  
 At the bifurcation, a higher radial mode of the same azimuthal
 wave grows to a finite amplitude while the first radial
 mode appears saturated around $Ta\sim10^6$ and even decays. 
 In terms of the physical quantities, this development results
 in strong velocity and temperature gradients near the boundaries
 and the concentration of the radial transport of heat and momentum
 in narrow plumes or jets. At the highest values of Taylor numbers explored,
 $2$ to $5\times 10^6$, that flow then developed temporal fluctuations.

\begin{figure*}
(a)\hspace{0.3\textwidth}(b)\hspace{0.3\textwidth}(c) \hfill ~ \\
\includegraphics[width=0.29\textwidth]{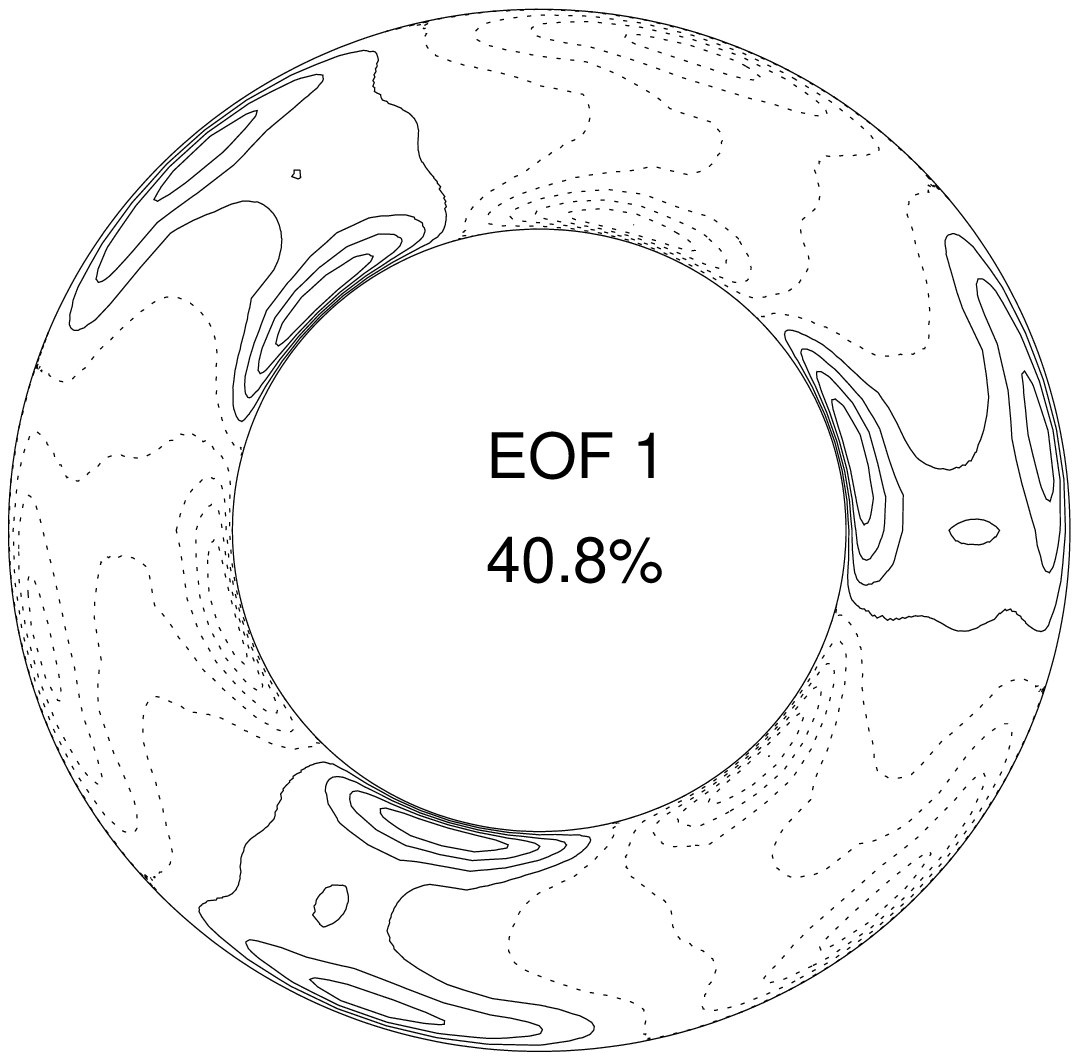}
\includegraphics[width=0.29\textwidth]{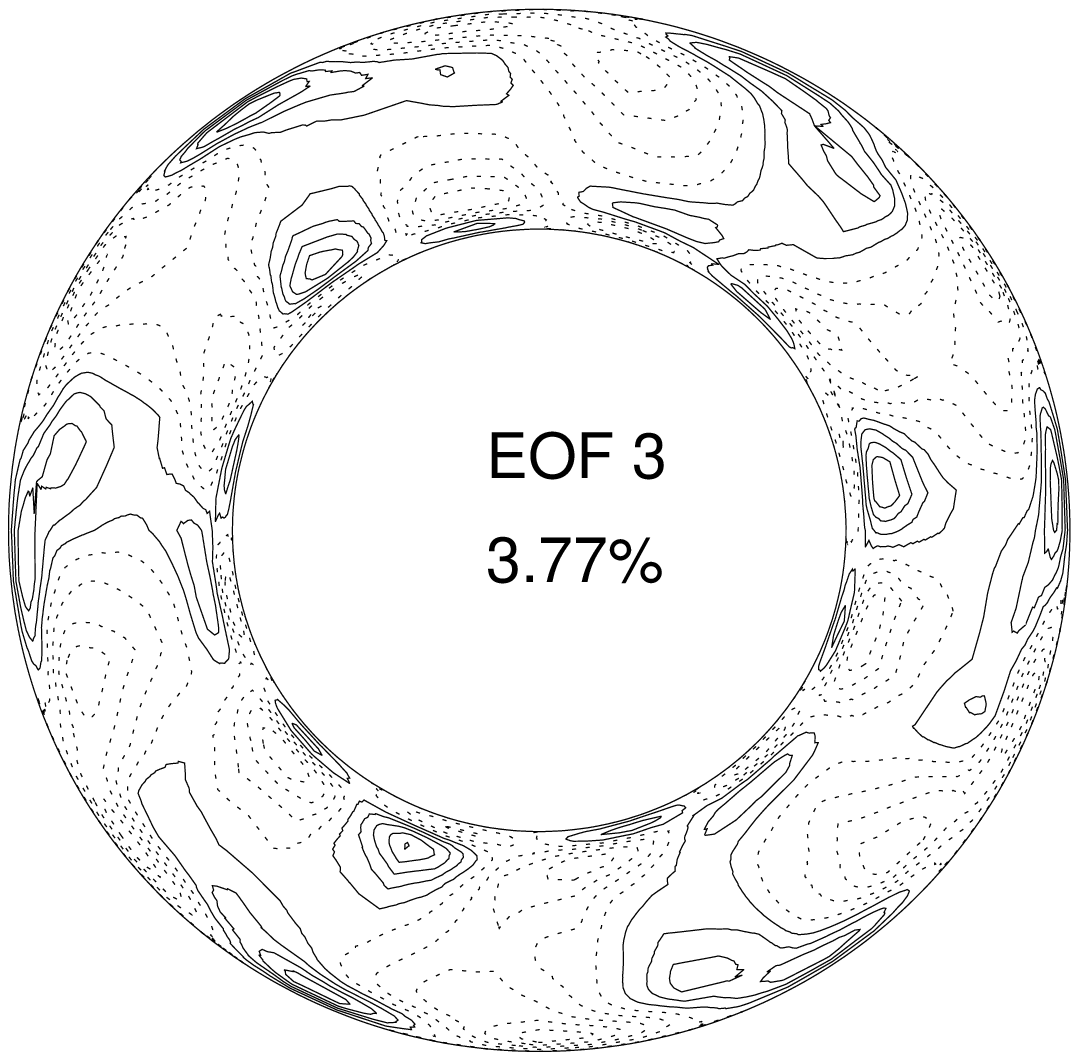}
\includegraphics[width=0.29\textwidth]{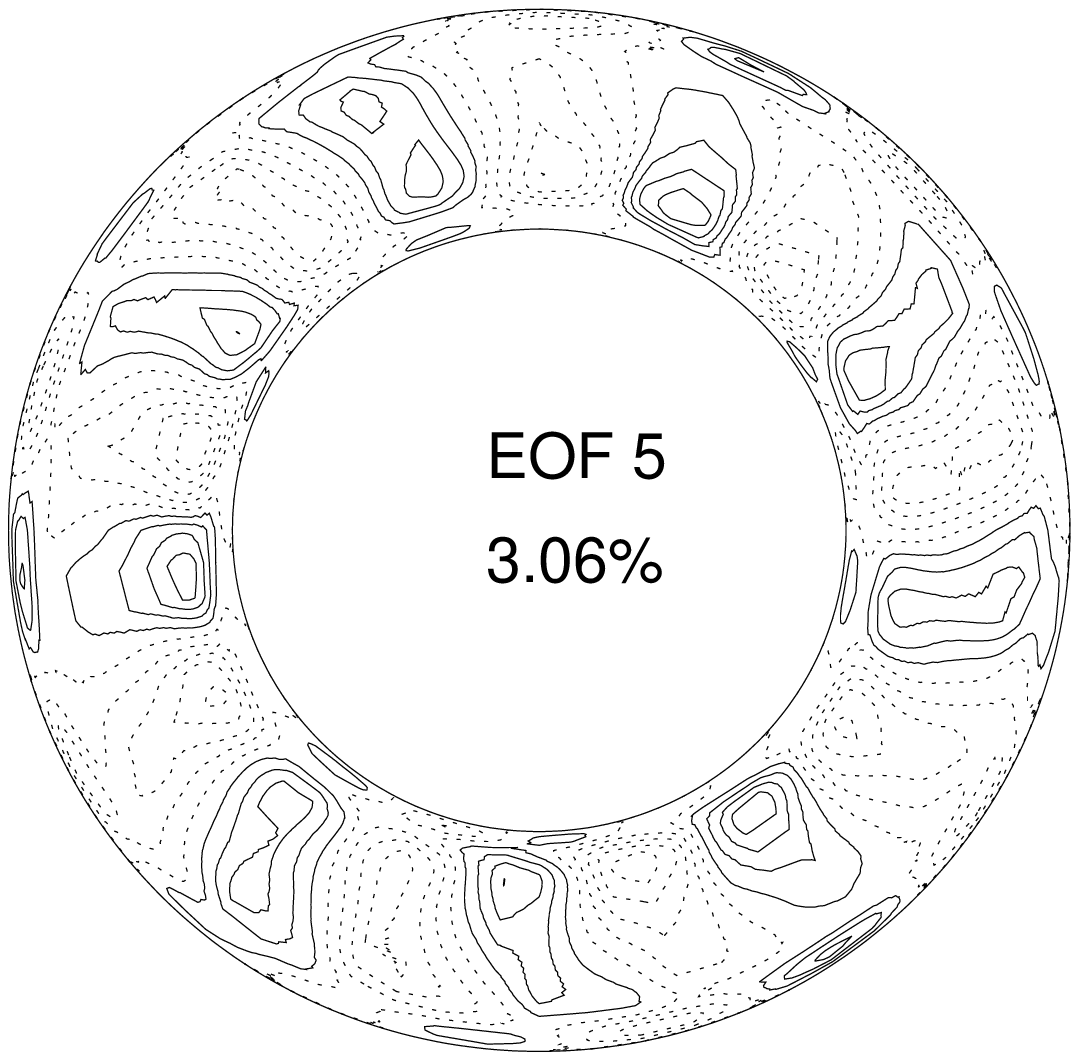}
\hfill \\
(d)\hspace{0.3\textwidth}(e)\hspace{0.3\textwidth}(f) \hfill ~ \\
\includegraphics[width=0.29\textwidth]{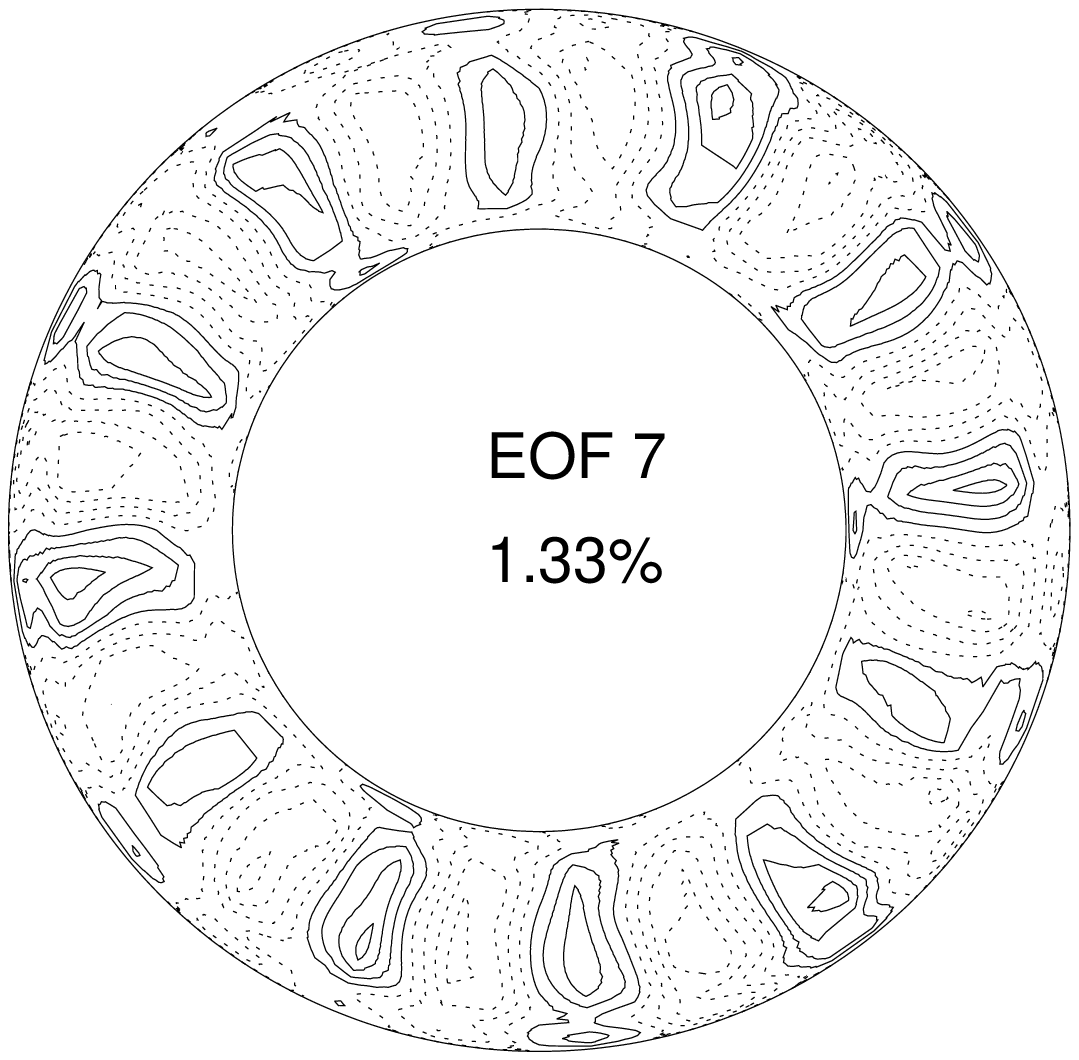}
\includegraphics[width=0.29\textwidth]{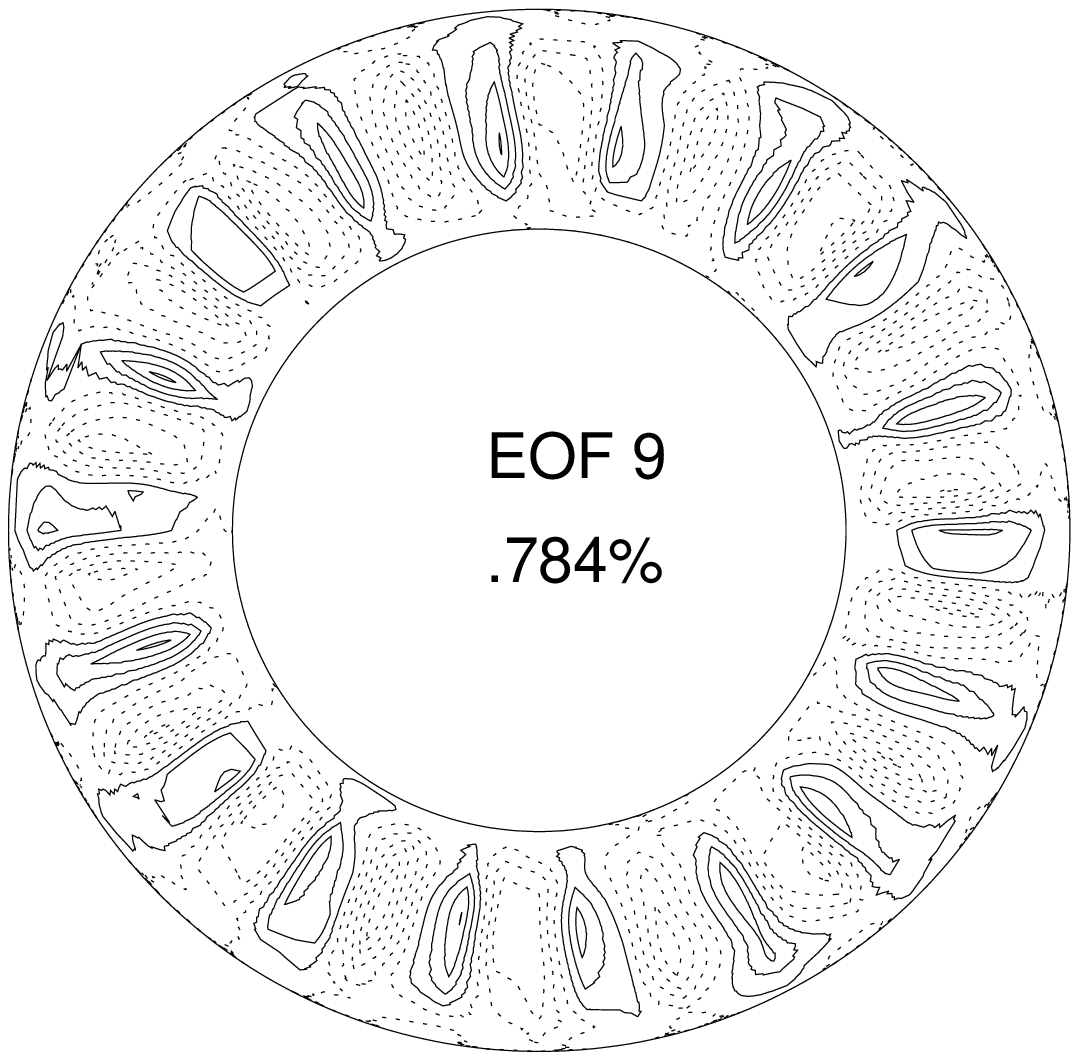}
\includegraphics[width=0.29\textwidth]{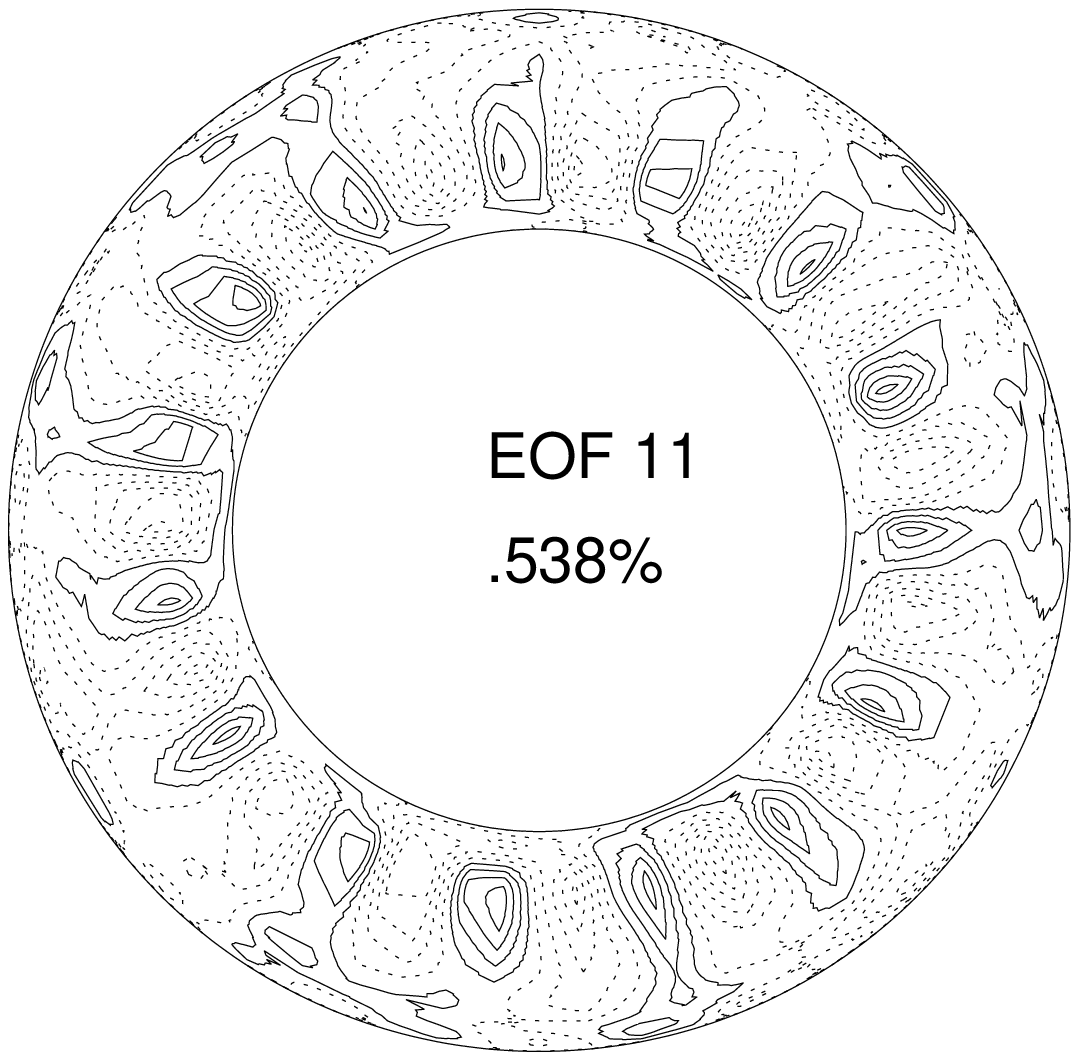}
\hfill
\caption{\label{SVeof3} Maps of the odd EOFs of the first six pairs EOFs of temperature at mid-height from
a simulation of baroclinic waves in air at Ta = $4.0 \times 10^6$. EOFs are shown
in order of decreasing contribution to the total temperature variance at this level.}
\end{figure*}

This change from a regular steady wave dominated by baroclinic instability to
 the transition zone characterized by small-scale fluctuations
 can be seen from the variations of eddy variables at two different values of
 rotation rates in the
azimuth-height maps at mid-radius in Figure~\ref{TZfields1}. 
 At $Ta= 0.235 \times 10^6$, the solution
 shows a pressure field which tilts westward with height and
 the temperature and vertical velocity fields which tilt eastward in such a
 way that the strongest upwelling is located around the strongest temperature
 gradient and strongest pressure gradient near the lower boundary. This behaviour is
 that of a typical baroclinic steady wave. 
 At higher rotation rate $Ta= 4. \times 10^6$, 
  the pressure field clearly shows very little
 phase tilt with height while the temperature field exhibits highly
 concentrated plumes of hot and cold air which have very little
 slope with height, interspersed with broad regions with relatively
 very weak horizontal thermal gradients. The thermal plumes are
 aligned azimuthally with the regions of strong azimuthal pressure gradient,
 consistent with optimising the correlation between radial (geostrophic)
 velocity and temperature perturbations.  Similarly, the regions of
 strong upward vertical velocity are also concentrated close to the
 strongest positive temperature anomalies, and {\em vice versa},
 consistent with an optimisation of $\overline{w'T'}$ 
 (where $w'$ and $T'$ represent departures in vertical velocity
 and temperature from their azimuthal mean values).  Eventually,
 the decrease of the tilt results in a virtually vertical,
 'barotropic' structure. Regions of
 strong downward motion (necessary to satisfy mass conservation)
 are concentrated in plumes or jets adjacent to the strong upwelling
 jets, forming `cross-frontal' circulations which show  some similarities
 with those inferred for atmospheric frontal regions in developing
 cyclones (\cite{Hoskins82}). The overall impression is that the flow in this
 region of parameter space is strongly nonlinear and much modified
 from the simple, linearly unstable Eady solution found at much lower Taylor number.
        
\section{The SV characteristics}

The spatial structure associated especially with the onset of structural
 vacillation is of particular interest, since it may be associated with
 bifurcations involving quite different spatial modes. As a means of isolating
 the dominant patterns in the flow, we have made use of a form of Empirical
 Orthogonal Function (EOF) analysis (\cite{Preisendorfer88}), in which time sequences
 of spatial maps e.g. of temperature were analysed to obtain the covariance
 matrix of every spatial point with every other in the two-dimensional field.

 Figures~\ref{SVeof3} show the first
six pairs of EOFs at $Ta = 4.0\times 10^6$, which together account for a little over 98\%
of the temperature variance. However, the first six EOFs appear as three
conjugate pairs and represent 93.2\% of the variance,
 indicating a slightly broader spread into the higher order patterns.
 The first two EOFs, e.g.~Fig.~\ref{SVeof3}(a), correspond to pattern
 dominated by $m=3$, where the radial
 structure seems somewhat more concentrated towards the side boundaries,
 with only weak amplitude throughout most of the interior radii.
 EOFs 3 and 4, Fig.~\ref{SVeof3}(b), show a pattern dominated by $m = 6$, but with even
more contorted phase shifts with radius. The overall impression is of a pattern represented by
$m = 6$ in azimuth and $l = 4$ in radius (where $l$ is a nondimensional radial wavenumber).
Subsequent EOFs display patterns with a reasonably simple $l = 1$ radial structure (except
for (c), which also shows evidence for $l = 4$ dependence on $r$ in EOFs 3 and 4), and increasing
azimuthal harmonics of $m = 3$.  The higher order EOFs in Figs~\ref{SVeof3}(c) and (d) generally have
 a much simpler radial structure, with a simple maximum in amplitude near
 mid-radius and only weak phase-tilts with radius. They again appear in
 conjugate pairs, and represent other azimuthal harmonics of $m=3$. The complex
 phase variations with radius in EOFs 3 and 4 demonstrate
 the growth of higher radial modes as previously
supposed in Figure~\ref{Bifurcations}, though the patterns in (c) and (d) of
Figs~\ref{SVeof3} would seem to suggest such anti-correlations ought
 to be strongest for $m = 6$ rather than the dominant $m=3$.

\begin{figure}
(a) \includegraphics[width=0.28\textwidth]{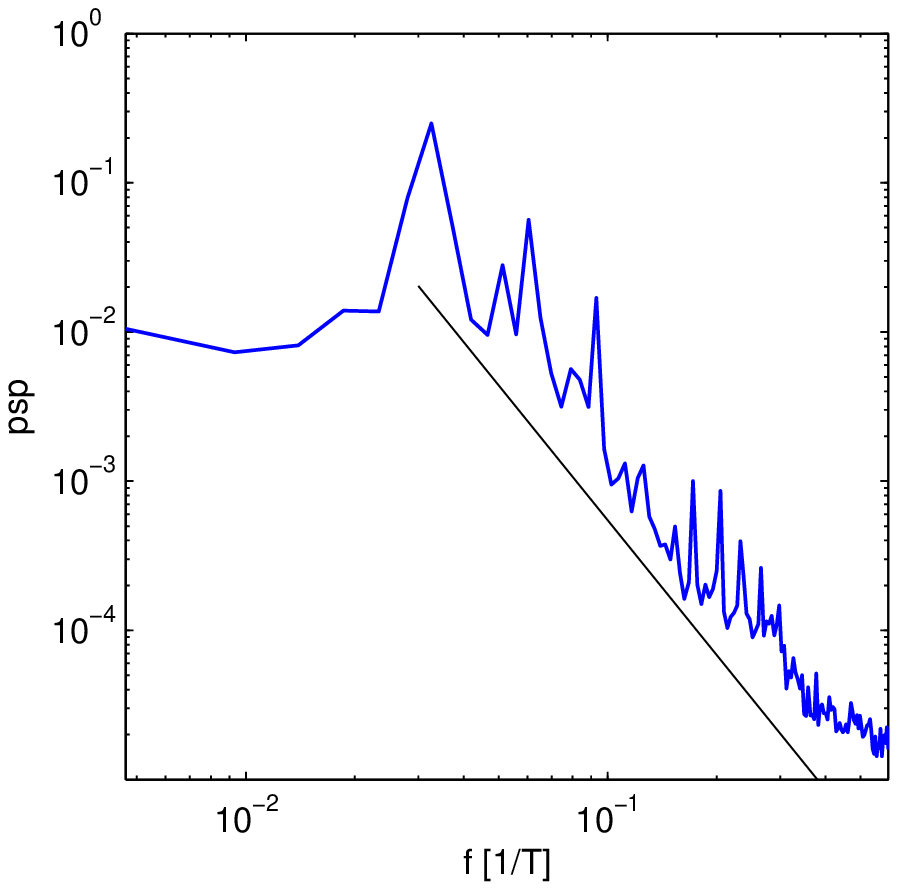}
(b) \includegraphics[width=0.28\textwidth]{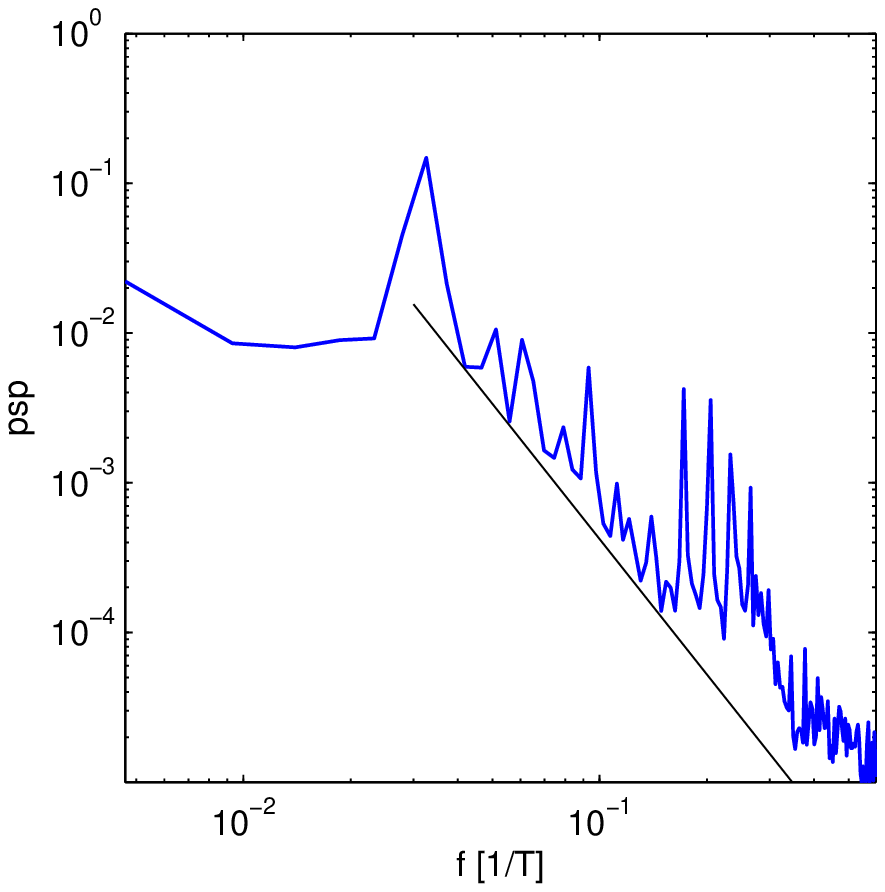}
(c) \includegraphics[width=0.28\textwidth]{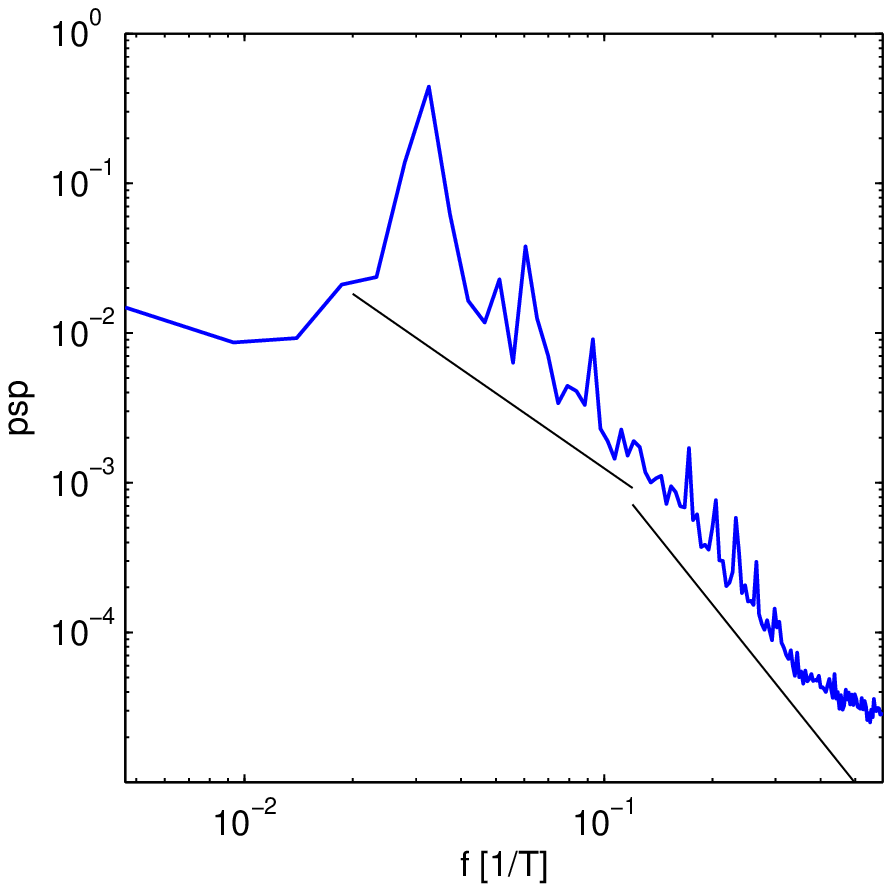}
\caption{\label{Spectra} Power spectra from the time series of the temperature
  after removing the steady wave: (a) near the inner wall, (b) at mid-radius, and (c) near the outer wall.  All show a line $\sim f^{-3}$ but (c) also shows a line $\sim f^{-5/3}$ at the lower frequencies.}
\end{figure}

Spatially averaged power spectra from the temperature residuals, after
 removing the steady wave, at mid-height
 and three radial positions, at mid-radius and around 15\% from each wall, 
  are shown in Figure~\ref{Spectra}.  
 The main frequency ($\sim 0.03 T^{-1}$, with the time scale $T=1/2\Omega$) is that of the emission
 of the perturbations described by EOFs 3 and 4.  The spectrum falls off at  all radial positions,
 with some distinctive peaks over the general decay. 
 While the decay at higher frequencies is largely consistent
 with a $f^{-3}$ law at all postions, a slight flattening in the
 frequency range between that of the main perturbation and about
 $0.1T^{-1}$ can be observed near the outer wall.
  The behaviour of the spectrum in that range appears to be closer to a  $f^{-5/3}$ law.  

The spectral evidence suggests that the flow investigated here consists
 of a fairly steady large-scale convection pattern in the form of
 three pairs of hot and cold radial jets.  From those jets, smaller
 perturbations are emitted at relatively regular intervals where a
 jet approaches a wall.   The overall flow field responds in a cascade
 of faster (and presumably smaller) fluctuations which appear consistent
 with two-dimensional, quasi-geostrophic turbulence over a wide range
 of frequencies similar to results of DNS of geostrophic turbulence. 
 Recently, similar energy spectra were observed for geostrophic turbulence in
 a square box (\cite{Lindborg00,Waite06}).     

\section{Conclusions}
In the present work, we find the emergence of a new steady wave  
solution with complex radial structure, associated with a change in  
the balance of forces to a centrifugally-dominated regime, followed  
(at high Taylor number) by a further transition to a form of  
structural vacillation associated with regular m = 3 eddies driven by  
centrifugal buoyancy. The computations have been carried out on the NEC SX-5
of the IDRIS (CNRS, Orsay, France). The authors are grateful to the British
Council and the CNRS for funding the collaboration between the French and UK
partners in a joint programme Alliance.



\begin{thebibliography}{}

\bibitem[Fowlis \& Hide (1965)]{Fowlis65}
{W. W. Fowlis and R. Hide} 1965 
{Thermal convection in a rotating annulus of liquid : effect of viscosity on
 the transition between axisymmetric and non-axisymmetric flow regimes.}
 \textit{J. Atmos. Sci.}
 \textbf{22} 541-558

\bibitem[Fr{\"u}h \& Read (1997)]{Fruh97}
{W.-G. Fr{\"u}h and P. L. Read} 1997 
{Wave interactions and the transition to chaos of baroclinic waves in a thermally driven rotating annulus.}
 \textit{Phil.~Trans.~R.~Soc.~Lond.~(A).}
 \textbf{355} 101-153

\bibitem[Hignett \etal (1985)]{Hignett85}
{P. Hignett, A. A. White, R. D. Carter, W. D. N. Jackson and R. M. Small} 1985
{A comparison of laboratory measurements and numerical simulations of baroclinic wave flows in a rotating cylindrical annulus.}
\textit{Quart. J. R. Met. Soc.}
\textbf{111} 131-154

\bibitem[Hoskins (1982)]{Hoskins82}
{B. J. Hoskins} 1982 
{The mathematical theory of frontogenesis.}
\textit{Ann. Rev. Fluid Mech.}
\textbf{14} 131-154

\bibitem[Lindborg \& Alvelius (2000)]{Lindborg00}
{E. Lindborg and K. Alvelius} 2000
{The kinetic energy spectrum of the two-dimensional enstrophy turbulence cascade.}
\textit{Phys. Fluids.}
\textbf{12(5)} 945-947

\bibitem[Pierrehumbert \& Swanson (1995)]{Pierrehumbert95}
{R. T. Pierrehumbert and K. L. Swanson} 1995
{Baroclinic instability.}
\textit{Ann. Rev. Fluid Mech.}
\textbf{27} 419-467

\bibitem[Preisendorfer (1988)]{Preisendorfer88}
{R. W. Preisendorfer} 1988
\textit{Principal Component Analysis in Meteorology and Oceanography.}
{Elsevier, Amsterdam}

\bibitem[Randriamampianina \etal (2006)]{Randria06}
{A. Randriamampianina, W.-G. Fr\"uh, P. Maubert and P. L. Read} 2006
{Direct Numerical Simulation of bifurcations in an air-filled rotating baroclinic annulus.}
\textit{J. Fluid Mech.}
\textbf{561} 359-389 

\bibitem[Read (2001)]{Read01}
{P. L. Read} 2001
{Transition to geostrophic turbulence in the laboratory, and as a paradigm in atmospheres and oceans.}
 \textit{Surveys in Geophys.}
 \textbf{22} 265-317

\bibitem[Read \etal (1998)]{Read98}
{P. L. Read and M. Collins and W.-G. Fr{\"u}h and S. R. Lewis and A. F. Lovegrove} 1998
{Wave interactions and baroclinic chaos: a paradigm for long timescale variability in planetary atmospheres.}
\textit{Chaos, Solitons \& Fractals.}
\textbf{9} 231-249

\bibitem[Waite \& Bartello (2006)]{Waite06}
{M.L. Waite and P. Bartello} 2006
{The transition from geostrophic to stratified turbulence.}
\textit{J. Fluid Mech.}
\textbf{568} 89-108

\bibitem[Weng \& Barcilon (1987)]{Weng87}
{H.-Y. Weng and A. Barcilon} 1987 
{Wave structure and evolution in baroclinic flow regimes.}
\textit{Quart. J. R. Met. Soc.}
\textbf{113} 1271-1294
  
\end{thebibliography}
 \end{document}